\documentclass[3p,times,twocolumn]{elsarticle}
\usepackage{amsmath}
\usepackage{amssymb}
\usepackage{graphicx}
\usepackage{slashed} 

\usepackage{ecrc}

\volume{00}

\firstpage{1}

\journalname{Nuclear Physics B Proceedings Supplement}

\runauth{}

\jid{nuphbp}

\jnltitlelogo{Nuclear Physics B Proceedings Supplement}

\usepackage[figuresright]{rotating}

\newcommand{\GeV}{\mbox{ GeV}}

\newcommand{\cm}{\mbox{ cm}}

\begin{document}

\begin{frontmatter}



\dochead{}

\title{ Direct and indirect constraints on isospin-violating dark matter}


\author{Yu-Feng Zhou}

\address{
 State Key Laboratory of Theoretical Physics,\\
  Kavli Institute for Theoretical Physics China,\\
  Institute of  Theoretical Physics, Chinese Academy of Sciences,\\
  Beijing, 100190, P.R. China
}

\begin{abstract}
  The scenario of isospin-violating dark matter (IVDM) with destructive
  interference between DM-proton and DM-neutron scatterings provides a
  potential possibility to reconcile the experimental results of DAMA, CoGeNT
  and XENON. We explore the constraints on the IVDM from other direct
  detection experiments such as CRESST and SIMPLE, etc. and from the indirect
  DM searches such as the antiproton flux measured by BESS-Polar II.  The results
  show that the relevant couplings in IVDM scenario are severely constrained.
  
\end{abstract}

\begin{keyword}
dark matter \sep isospin-violating interaction \sep cosmic-ray antiproton

\end{keyword}

\end{frontmatter}





Some of the recent dark matter (DM) direct detection experiments such as DAMA
\cite{0804.2741, 0808.3607,1002.1028}, CoGeNT \cite{1002.4703,1106.0650} and
CRESST-II \cite{1109.0702} have reported events which cannot be explained
by  conventional  backgrounds. The excesses, if interpreted in terms of DM particle elastic
scattering off target nuclei, may imply light DM particles with mass around
8-10 GeV and scattering cross section around $10^{-40}\cm^{2}$. Other
experiments such as CDMS-II \cite{1010.4290, 1011.2482}, XENON10/100
\cite{1104.3088,1104.2549}, and SIMPLE \cite{1106.3014} etc., have reported
null results in the same DM mass range. 

A commonly adopted assumption on interpreting the DM direct detection data is
that in spin-independent scatterings the DM particle couplings to proton
$(f_{p})$ and to neutron $(f_{n})$ are nearly the same, i.e. $f_{n}\approx
f_{p}$, which makes it straight forward to extract the DM-nucleon scattering
cross sections. It is a good approximation for neutralino DM and DM models
with Higgs portal, e.g, the scalar DM in left-right models\cite{Guo:2011zze,Guo:2010sy,Guo:2010vy,Guo:2008si} and 4th generation Majorana neutrino DM \cite{Zhou:2011fr}.
However, in generic
cases, the interactions may be isospin-violating
~\cite{hep-ph/0307185,hep-ph/0504157,1003.0014,1004.0697,1102.4331,1105.3734,1112.6364}.
In this scenario, the DM particle couples to
proton and neutron with different strengths, possible destructive interference
between the two couplings can weaken the bounds of  XENON10/100 and move the
signal regions of DAMA and CoGeNT to be closer to each other
\cite{1102.4331,1105.3734}. In order to reconcile the data of DAMA, CoGeNT and
XENON10/100, a large destructive interference corresponding to $f_{n}/f_{p}\approx -0.7$ is
required \cite{1102.4331}.

Possible constraints on IVDM from the cosmic neutrinos and gamma ray on IVDM have been
discussed previously  in Refs. \cite{ 1103.3270,1106.4044,1112.4849}. Recently the
BESS-Polar II experiment has measured the antiproton flux in the energy range
from 0.2 GeV to 3.5 GeV \cite{1107.6000} which have  higher precision compared
with that from PAMELA  \cite{1103.2880} at low energies.
In this talk, we discuss on the direct and indirect constraints on IVDM with focus
on the cosmic-ray antiproton constraints. The details of our analysis can be found in
Ref. \cite{Jin:2012jn}.

For a DM particle $\chi$ elastically scattering off a
target nucleus,  the
differential scattering cross section can be written as
\begin{align}\label{eq:9}
  \frac{d\sigma}{dE_{R}}=
  \frac{m_{A}F^{2}(E_{R})}{2\mu_{A}^{2} v^{2}}\sigma_{0}  \ ,
\end{align}
where $F(E_{R})$ is the form factor of the nucleon and
$\mu_{A}=(m_{\chi}m_{A})/(m_{\chi}+m_{A})$ is the DM-nucleus reduced mass. 
 The quantity $\sigma_{0}$ can be understood as
the total scattering cross section at the limit of zero-momentum transfer which
is related to $f_{p(n)}$  through
\begin{align}\label{eq:21}
\sigma_{0}=\frac{\mu_{A}^{2}}{\pi}\left[Zf_{p} +(A-Z)f_{n}\right]^{2} \ ,
\end{align}
where  $Z$ is the  atomic number and $A$ is the atomic mass number.
Under the assumption that the scattering is isospin conserving (IC),
i.e., $f_{n}\approx f_{p}$, the total cross section $\sigma_{0}$ is independent of
$Z$ and only  proportional to $A^{2}$.  One can define a
cross section $\sigma_{p}^{IC}$ which is the value of $\sigma_{p}$ extracted  from
$\sigma_{0}$ under the assumption of IC interaction as
\begin{align}\label{eq:19}
\sigma_p^{IC} \equiv \frac{\mu_{p}^{2}}{\mu_{A}^{2}A^{2}} \sigma_{0} \ .
\end{align}
In the generic case where $f_{n} \neq f_{p}$, the true value of $\sigma_{p}$ will
differ from $\sigma^{IC}_{p}$ by a factor $F(f_n/f_p)$ which depends on the
ratio $f_n/f_p$ and the target material
\begin{align}\label{eq:19}
\sigma_p = F(f_n/f_p) \sigma_p^{IC} .
\end{align}
If  the target material consists of  $N$ kind of relevant
nuclei with atomic numbers $Z_{\alpha} \ (\alpha=1,\dots, N)$ and
fractional number abundances  $\kappa_{a}$,  and for each nucleus $Z_{\alpha}$
there exists  $M$ type of isotopes found in nature with atomic mass number
$A_{\alpha i}$ and fractional number abundance $\eta_{\alpha i}
\ (i=1,\dots,M)$, the expression of $F(f_{n}/f_{p})$ can be explicitly written as
\begin{align}
\label{eq:34}
F(f_{n}/f_{p})=
\frac{\sum_{\alpha,i}\kappa_{\alpha}\eta_{\alpha i} \mu_{A_{\alpha i}}^{2}A_{\alpha i}^{2}}{\sum_{\alpha, i}\kappa_{\alpha}\eta_{\alpha i}\mu_{A_{\alpha i}}^{2}[Z_{\alpha}+(A_{\alpha i}-Z_{\alpha})f_{n}/f_{p}]^{2}} \ ,
\end{align}
where $\mu_{A_{\alpha i}}$ is the reduced mass for the DM and the nucleus  with
atomic mass number $A_{\alpha i}$. For a given target material $T$, there is a particular value of $f_{n}/f_{p}$ which corresponds
to the maximal possible value of $F(f_n/f_p)$
\begin{align}
\label{eq:44}
\xi_{T}\equiv
=-\frac{\sum_{\alpha,i} \kappa_{\alpha}\eta_{\alpha i} \mu_{A_{\alpha i}}^{2}(A_{\alpha i}-Z_{\alpha})Z_{\alpha}}{\sum_{\alpha, i} \kappa_{\alpha}\eta_{\alpha i} \mu_{A_{\alpha i}}^{2}(A_{\alpha i}-Z_{\alpha})^{2}} \ .
\end{align}
The value of $\xi_{T}$ varies  with
target material. In Tab. \ref{tab:xi}, we list the values of
$\xi_{T}$ for some typical material utilized by the current or future experiments.

\begin{table*}
  \begin{center}
    \begin{tabular}{ccccccccc}\hline\hline
      & Xe     &Ge     & Si  &Na(I)     &Ca(W)$\mbox{O}_{4}$  & $\mbox{C}_{2}\mbox{ClF}_{5}$ & CsI  & Ar\\
      \hline
      $\xi_{T}$& -0.70  &-0.79 & -1.0 &-0.92(-0.73) &-1.0(-0.69)                                     &-0.92                                                &-0.71 &-0.82\\
      \hline\hline
    \end{tabular}
    \caption{Values of $\xi_{T}$ for different target material. For NaI, the two values -0.92 and -0.73 correspond to the scattering off
      Na and NaI respectively. Similarly, for $\text{CaWO}_{4}$, the two values  -1.0 and  -0.69 corresponds to the scattering without and
    with tungsten nuclei respectively.}  \label{tab:xi}
  \end{center}
\end{table*}

\begin{figure}[htb]
\begin{center}
   \includegraphics[width=0.9\columnwidth]{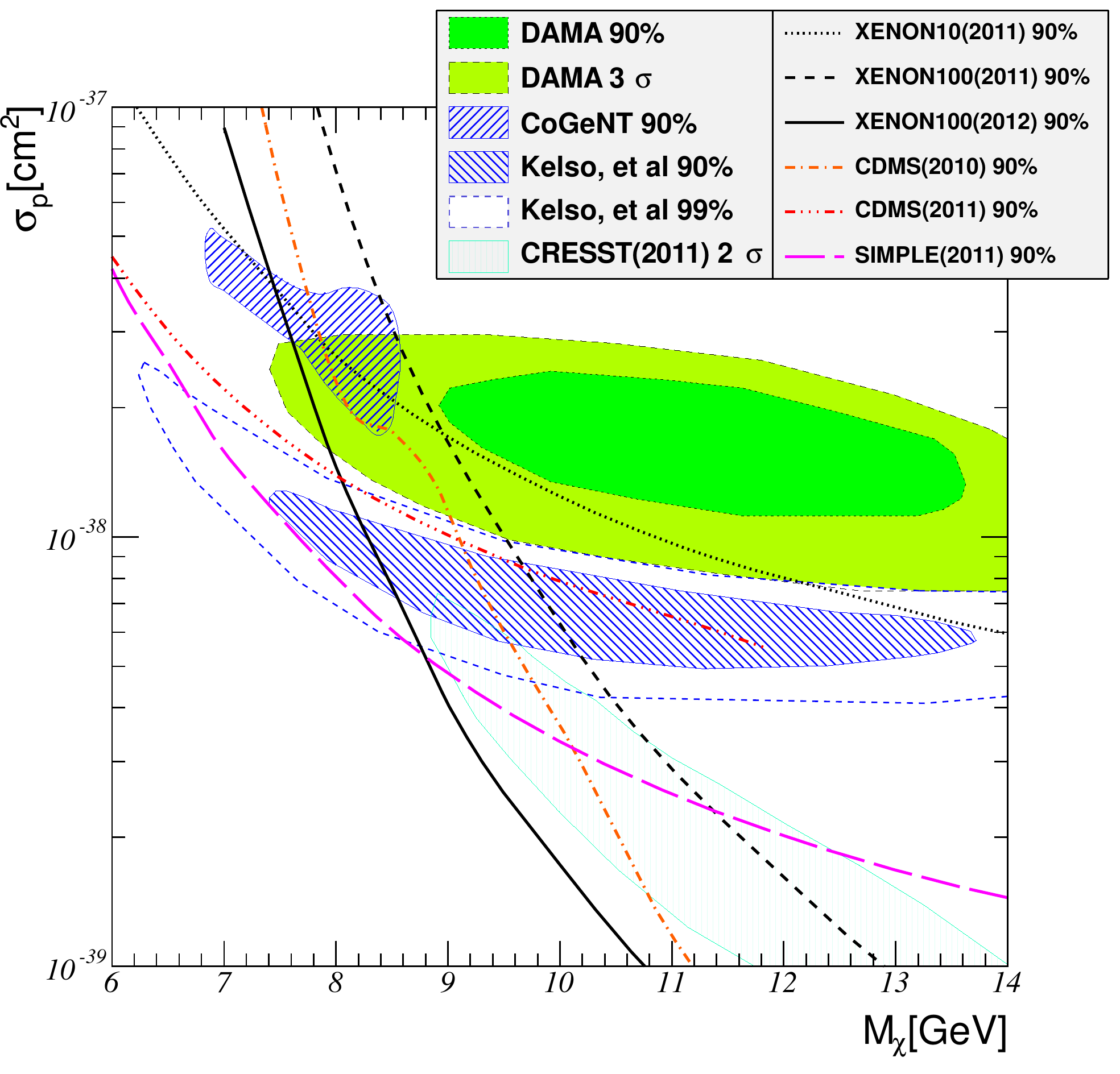}
\end{center}
\caption{Favored regions and limits in the $(\sigma_p, m_{\chi})$ plane from
  various experiments for $f_{n}/f_{p}=-0.70$ such as DAMA \cite{0808.3607},
  GoGeNT unmodulated data \cite{1106.0650}, CoGeNT unmodulated data with
  surface event rejection factors taken from Kelso, etal \cite{1110.5338},
  CRESST-II \cite{1109.0702}, XENON10/100 \cite{1104.3088,1207.5988}, CDMS
  \cite{1010.4290,1011.2482} and SIMPLE \cite{1106.3014}.  }
\label{fig:direct_IV}
\end{figure}

  If the $\xi_{T}$ values of the target material used by two experiments are
very close to each other, the tension between the two experimental results, if
exists, is less affected by the effect of isospin violation. From
Tab. \ref{tab:xi} one finds that $\xi_{\text{Na}}\approx
\xi_{\text{C}_2\text{ClF}_5}=-0.92$, $\xi_{\text{Xe}}\approx
\xi_{\text{CsI}}\approx -0.7$ and $\xi_{\text{Si}}\approx
\xi_{\text{Ca(W)O}_{4}}=-1.0$.
Thus the tension between DAMA
signal from Na recoil and the upper bound from SIMPLE is unlikely to be alleviated by
isospin violation, which can be clearly seen in Fig.~\ref{fig:direct_IV}.  Similarly, if there exists
contradictions  between XENON and KIMS, CoGeNT and the Ar based experiments
such as  DarkSide, it  can hardly be  explained by isospin violating scattering.
 The SIMPLE result is also useful in
comparing with the CRESST-II which utilizes Ca(W)$\mbox{O}_{4}$ which has
$\xi_{\text{Ca(W)O}_{4}}=-1.0$.
Obviously, for the experiments use the same target material, the possible
tension between them cannot be relaxed by isospin violation, such as the
tension between CoGeNT and CDMS-II, as both use germanium  as target nucleus.

In Fig. \ref{fig:direct_IV}, the allowed regions by the current experiments
are shown in the $(\sigma_{p},m_{\chi})$ plane for $f_{n}/f_{p}=-0.70$ . It
can be seen that the overlapping region between GoGeNT and DAMA may still be
consistent with the exclusion curve from the XENON100 2011 data
\cite{1104.2549}. However, If one considers the recently updated upper bounds
from XENON100 \cite{1207.5988}, the main bulk of the overlapping region is
excluded for both the GoGeNT results with and without surface event rejection
corrections, which challenges the IVDM as a scenario to reconcile the results
of DAMA, CoGeNT and XENON. The overlapping region between DAMA and CoGeNT
seems also to be excluded by the results of SIMPLE \cite{1106.3014} and
CDMS-II independently \cite{1010.4290,1011.2482}.  Note however that there
still exists controversies regarding the detector stability of SIMPLE
experiments \cite{1106.3559,1107.1515}, the recoil energy calibration of CDMS
experiment \cite{1103.3481} and the extrapolation of the measured
scintillation efficiency to lower recoil energy in the previous XENON100 data
analysis \cite{1005.0838,1005.2615}.

We assume that the DM particles interact with the SM light quarks through some
heavy mediator particles much heavier than the DM particle
such that both the scattering and the annihilation processes can be  effectively
described by a set of high dimensional contact operators
\begin{equation}\label{eq:36}
\mathcal{L}=\sum_{i,q} a_{iq}\mathcal{O}_{iq} \ .
\end{equation}
If the DM particles are Dirac fermions, the  relevant operators arising  from  scalar  or pseudoscalar  interactions are given by
\begin{eqnarray}\label{eq:1}
  \mathcal{O}_{1q} &= &\bar{\chi}\chi\bar{q}q,
  \mathcal{O}_{2q} =\bar{\chi}\gamma^{5}\chi\bar{q}q, 
  \nonumber\\
\mathcal{O}_{3q}&=&\bar{\chi}\chi\bar{q}\gamma^{5}q, 
\mathcal{O}_{4q}=\bar{\chi}\gamma^{5}\chi\bar{q}\gamma^{5}q  .
\end{eqnarray}
The operators from vector or axial-vector  type interactions are
\begin{eqnarray}\label{eq:2}
\mathcal{O}_{5q}&=& \bar{\chi}\gamma^{\mu}\chi\bar{q}\gamma_{\mu}q, \
\mathcal{O}_{6q}=\bar{\chi}\gamma^{\mu}\gamma^{5}\chi\bar{q}\gamma_{\mu}q, \
 \nonumber\\
\mathcal{O}_{7q}&=&\bar{\chi}\gamma^{\mu}\chi\bar{q}\gamma_{\mu}\gamma^{5}q, \
\mathcal{O}_{8q}=\bar{\chi}\gamma^{\mu}\gamma^{5}\chi\bar{q}\gamma_{\mu}\gamma^{5}q ,
\end{eqnarray}
and the ones from the tensor  interactions are
\begin{equation}\label{eq:3}
\mathcal{O}_{9q}=\bar{\chi}\sigma^{\mu\nu}\chi\bar{q}\sigma_{\mu\nu}q,  \quad
\mathcal{O}_{10q}=\bar{\chi}\sigma^{\mu\nu}\gamma^{5}\chi\bar{q}\sigma_{\mu\nu}q .
\end{equation}
If the DM particles are Majorana particles the vector and tensor operators are
vanishing identically. Among these operators only $\mathcal{O}_{1q}$ and
$\mathcal{O}_{5q}$ contribute to spin-independent scattering cross sections at
low velocities.  The scattering cross sections induced by the operators
$\mathcal{O}_{2q}$ and $\mathcal{O}_{6q}$ are velocity suppressed. The
operators $\mathcal{O}_{7q}$ and $\mathcal{O}_{8q}$ contribute only to
spin-dependent scattering cross section, and the nucleus matrix elements for
the operators $\mathcal{O}_{3q}$, $\mathcal{O}_{4q}$, $\mathcal{O}_{9q}$ and
$\mathcal{O}_{10q}$ are either vanishing or negligible.
The DM annihilation into quarks through $\mathcal{O}_{1q}$ is a $p$-wave
process, which is velocity suppressed. It does not contribute to the cosmic
antiproton flux, but  still contributes to  the DM relic density as $p$-wave processes
is non-negligible at freeze out.

Similarly, for DM being a complex scalar $\phi$,  possible operators are
\begin{eqnarray}\label{eq:5}
\mathcal{O}_{11}&=&2m_{\phi}(\phi^{*}\phi) \bar{q}q , \
\mathcal{O}_{12}=2m_{\phi}(\phi^{*}\phi) \bar{q} \gamma^{5}q , \
\nonumber\\
\mathcal{O}_{13}&=&(\phi^{*}\overleftrightarrow{\partial_{\mu}}\phi )\bar{q} \gamma^{\mu} q , \
\mathcal{O}_{14}=(\phi^{*}\overleftrightarrow{\partial_{\mu}}\phi )\bar{q} \gamma^{\mu} \gamma^{5}q .
\end{eqnarray}
Among those only $\mathcal{O}_{11}$ and $\mathcal{O}_{13}$ contribute to the
spin-independent scatterings. The DM annihilations through operator
$\mathcal{O}_{13}$ are $p$-wave processes.

In summary,
we only consider  four operators
$$
\mathcal{O}_{1q}, \  \mathcal{O}_{5q}, \  \mathcal{O}_{11q}, \mbox{ and } \mathcal{O}_{13q} ,
$$
which are relevant to IVDM.

The DM couplings to nucleons $f_{p,n}$ can be expressed in terms of  the DM couplings to quarks $a_{iq}$ as follows
\begin{equation}
\label{eq:37}
f_{p(n)}=\sum_{q} B^{p(n)}_{iq} a_{iq} .
\end{equation}
For the Dirac DM with scalar interaction $a_{1q} \bar{\chi}\chi \bar{q}q$, one has  $B^{p(n)}_{1q}=f^{p(n)}_{Tq}\ m_{p(n)}/m_{q} $ for $q=u,d,s$
and $B^{p(n)}_{1q}=(2/27)f^{p(n)}_{TG} m_{p(n)}/m_{q}$ for $q=c,b,t$,
where $f^{p(n)}_{Tq}$ is the DM coupling to light quarks obtained from the  $\sigma$-term $\left\langle N|m_{q}\bar{q}q|N\right\rangle =f_{Tq}^{N}M_{N}$,
and $
f^{p(n)}_{TG}=1-\sum_{q=u,d,s}f^{p(n)}_{Tq}$.  
In order to maximize the isospin violating effect, the coefficients
$B_{1s,1c,1b,1t}^{p,n}$ must be strongly suppressed.  Assuming that the
DM-nucleon couplings are dominated by the DM couplings to the first generation
quarks, the ratio $f_{n}/f_{p}$ is given by
\begin{equation}
\label{eq:40}
\frac{f_{n}}{f_{p}} \approx \frac{B^{n}_{1u}a_{1u}+B^{n}_{1d} a_{1d}}{B^{p}_{1u}a_{1u}+B^{p}_{1d} a_{1d}} .
\end{equation}
The value of  $f_{n}/f_{p}=-0.7$ can be translated into  $a_{1d}/a_{1u}= -0.93$ at quark level. This  value is the same for complex scalar DM.
For operator  $\mathcal{O}_{5q}$ one simply has
$B^{p(n)}_{5u}=2(1)$  and $B^{p(n)}_{5d}=1(2)$,  and  $B^{p(n)}_{q}=0$  for $q=c,s,t,b$.
\begin{equation}
\label{eq:41}
\frac{f_{n}}{f_{p}}=\frac{a_{5u}+2  a_{5d}}{2 a_{5u}+ a_{5d}} .
\end{equation}
Thus for $f_{n}/f_{p}=-0.7$, one finds $a_{5d}/a_{5u}=-0.89$.


Annihilation or decay of light DM particles in the galactic halo can
contribute to exotic primary sources of the low energy cosmic ray antiprotons,
which can be probed or constrained by the current satellite- and balloon-borne
experiments such as PAMELA and BESS-polar II, etc..  The predicted antiproton
flux from DM annihilation depends on models of the cosmic-ray transportation,
the distribution of Galactic gas, radiation field and magnetic field, etc.. It also
depends on the particle and nuclear interaction cross sections.
In this work, we use the numerical code GALPROP
\cite{astro-ph/9807150,astro-ph/0106567,astro-ph/0101068,astro-ph/0210480,astro-ph/0510335} which utilizes realistic astronomical
information on the distribution of interstellar gas and other data as input
and consider various kinds of data including primary and secondary nuclei,
electrons and positrons, $\gamma$-rays, synchrotron and radiation etc. in a
self-consistent way. 
In the GALPROP approach, we consider several diffusion models (parameter
configurations). The different results between the models can be regarded as an
estimate of theoretical uncertainties.

In the diffusion models of cosmic ray propagation, the Galactic halo where
diffusion occurs is parameterized by a cylinder with half height $Z_{h}$ and
radius $R=20-30$ kpc.  The densities of cosmic ray particles are vanishing at the
boundary of the halo. The processes of energy losses, reacceleration and
annihilation take place in the Galactic disc. The source terms for the
secondary cosmic rays are also confined within the  disc.
The diffusion  equation for the cosmic ray particle is given by
\begin{eqnarray}\label{eq:28}
  \frac{\partial \psi}{\partial t} &=&
  \nabla (D_{xx}\nabla \psi -\mathbf{V}_{c} \psi)
  +\frac{\partial}{\partial p}p^{2} D_{pp}\frac{\partial}{\partial p} \frac{1}{p^{2}}\psi
     \nonumber \\
  &&-\frac{\partial}{\partial p} \left[ \dot{p} \psi -\frac{p}{3}(\nabla\cdot \mathbf{V}_{c})\psi \right]
  \nonumber \\
  && -\frac{1}{\tau_{f}}\psi
  -\frac{1}{\tau_{r}}\psi
  +q(\mathbf{r},p)  ,
\end{eqnarray}
where $\psi(\mathbf{r},p,t)$ is the number density per unit of total particle
momentum which is related to the phase space density $f(\mathbf{r},p, t)$ as
$\psi(\mathbf{r},p,t)=4\pi p^{2}f(\mathbf{r},p,t) $. For steady-state diffusion, it is assumed that  $\partial  \psi/\partial t=0$.
The spatial diffusion coefficient $D_{xx}$ is parameterized as
\begin{equation}
\label{eq:29}
D_{xx}=\beta D_{0} \left( \frac{\rho}{\rho_{0}} \right)^{\delta}  ,
\end{equation}
where $\rho=p/(Ze)$ is the rigidity of cosmic ray particle and $\delta$ is the power spectral index which may  take
different values $\delta_{1}$ or $ \delta_{2}$ when $\rho$ is below or above the reference rigidity
$\rho_{0}$.  $D_{0}$ is a normalization constant, and $\beta=v/c$ is the velocity of the cosmic ray particle.
The convection term is related to the drift of antiproton from the Galactic
disc due to the Galactic wind.  The direction of the wind is usually assumed to be along the $z$-direction which is
perpendicular to the galactic disc and is a constant
$\mathbf{V}_{C}=[2\theta(z)-1] V_{c}$.
The  diffusion in momentum space is described by the reacceleration parameter $D_{pp}$ which is
related to the   Alfv$\grave{\mbox{e}}$n speed $V_{a}$ of the disturbances in the hydrodynamical plasma as
\begin{equation}
\label{eq:30}
D_{pp}=\frac{4V_{a}^{2} p^{2}}{3D_{xx}\delta(4-\delta^{2})(4-\delta)w} ,
\end{equation}
where $w$ stands for the level of turbulence which can be taken as $w=1$ as $D_{pp}$ depends only on the
 combination $V_{a}^{2}/w$.
 $\dot{p}$ is related to  the momentum loss rate,
$\tau_{f}$ and $\tau_{r}$ are the time scale for fragmentation and radioactive
decay respectively.


Thus the whole diffusion process depends on a number of parameters: $R$,
$Z_{h}$, $\rho_{0}$, $D_{0}$, $\delta_{1}/\delta_{2}$, $V_{c}$ , $V_{a}$,
$\rho_{s}$ and $\gamma_{1}/\gamma_{2}$. 
The recently updated analysis based on
Markov Chain Monte Carlo fits to the astrophysical  data show that $Z_{h}$ should be around
$4-7\text{ kpc}$ \cite{1001.0551,1011.0037}.  In order to obtain a
conservative upper bounds we choose $Z_{h}\approx 4 \text{ kpc}$ in our
analysis.

We consider several typical propagation models in GALPROP, and focus on the
models with the secondary antiproton background below  the current data,
which leaves room for DM contribution and results in conservative upper
bounds. The first one is the plain diffusion model (referred to as ``Plain'')
in which there is no reacceleration term \cite{astro-ph/0510335}. The second
one is the conventional model (referred to as ``Conventional'') with
reacceleration included \cite{astro-ph/0101068,astro-ph/0510335}. The last one
(referred to as ``Global-Fit'') is the model from a global fit to the relevant
astrophysical observables using Markov-Chain Monte Carlo method
\cite{1011.0037}. The main parameters of the three models are listed in
Tab. \ref{tab:prop-models}.
\begin{table*}
  \begin{center}
  \begin{tabular}{lllllllll}\hline\hline
    model &$R (\mbox{kpc})$& $Z_{h}(\mbox{kpc})$ & $D_{0}$ & $\rho_{0}$&$\delta_{1}/\delta_{2}$   & $V_{a}(\mbox{km}/\mbox{s})$ &$\rho_{s}$&$\gamma_{1}/\gamma_{2}$ \\
 \hline
 Plain                &30   &  4.0  & 2.2   &3  &    0/0.60       & 0    &40 &2.30/2.15\\
Conv.            &20   & 4.0   & 5.75 &4  & 0.34/0.34     & 36  & 9  & 1.82/2.36 \\
Fit        &20   & 3.9   & 6.59 &4  &  0.3/0.3        &39.2 &10 & 1.91/2.40\\
  \hline\hline
  \end{tabular}\end{center}
\caption{Propagation parameters in the ``Plain'' \cite{astro-ph/0510335}, ``Conventional'' \cite{astro-ph/0101068,astro-ph/0510335} and
  ``Global-Fit'' \cite{1011.0037} models used in the GALPROP code. $D_{0}$ is in units of  $10^{28}\mbox{cm}^{2}\cdot\mbox{s}^{-1}$, the break rigidities $\rho_{0}$ and $\rho_{s}$  are in units of GV.}
 \label{tab:prop-models}
\end{table*}

The primary source term from  the DM annihilation has the form
\begin{equation}\label{eq:ann-source}
q(\mathbf{r})=\eta n(\mathbf{r})^{2}\langle \sigma v_{\text{rel}}\rangle \frac{dN}{dp} ,
\end{equation}
where $n(\mathbf{r})=\rho(\mathbf{r})/m_{\chi}$, $\eta=1/2(1/4)$ if the DM
particle is (not) its own antiparticle and $dN/dp$ is the injection spectrum
per DM annihilation.
For the DM profile we took the isothermal profile
\cite{astro-ph/9712318}
\begin{equation}
\label{eq:32}
\rho(\mathbf{r})=\rho_{0}\left( \frac{r_{\odot}^{2}+R_{s}^{2}}{r^{2}+R_{s}^{2}} \right) ,
\end{equation}
where $r_{\odot}=8.5\text{ kpc}$ is the distance of Solar system from the
galactic center, $R_{s}=2.8\text{ kpc}$ and the local density is taken to be
$\rho_{0}=0.3\GeV\cm^{-3}$.  The choice of isothermal profile and the local
density is again to achieve a conservative estimate of the antiproton flux, if
the NFW profile is used, the predicted flux will be enhanced roughly by at most
$70\%$, thus more severe constraints are expected.



In the case where the DM particles are  thermal relics, possible large couplings to light quarks may
under predict the  DM relic abundance in comparison with the observed  value $\Omega
h^{2}=0.113\pm0.004$ \cite{1001.4538}. Thus the relic density can also impose upper
bounds on the relevant couplings.  Whereas the $p$-wave annihilation processes
give no contribution to the antiproton flux, in the calculation of DM relic
density, the $p$-wave annihilation is nonnegligible  as  the
typical relative velocity  of DM particles is $v_{\text{rel}}/c \approx
\mathcal{O}(0.3)$  at freeze out. The annihilation cross section times the relative velocity
can be expressed  as $\sigma v_{\text{rel}}=a +b v_{\text{rel}}^{2}$, where $a$ and $b$ are coefficients
corresponding to the $s$-wave and $p$-wave contributions. 
For light DM around 10 GeV, $g_{*}=61.75$, we find $x_{f}\approx 22$.  The relic
abundance is approximately  given by
\begin{equation}
\label{eq:46}
\Omega h^{2} \simeq \frac{1.07\times 10^{9} \mbox{ GeV}^{-1}}{M_{pl}}\frac{x_{f}}{\sqrt{g_{*}}}
\frac{1}{a+3b x_{f}} .
\end{equation}

We perform $\chi^{2}$ analysis to obtain upper bounds on the DM isospin
violating couplings to light quarks. 
For the sake of simplicity, we consider one operator at a time and ignore
interference between the operators.

\begin{figure}[htb]
\begin{center}
    \includegraphics[width=0.9\columnwidth]{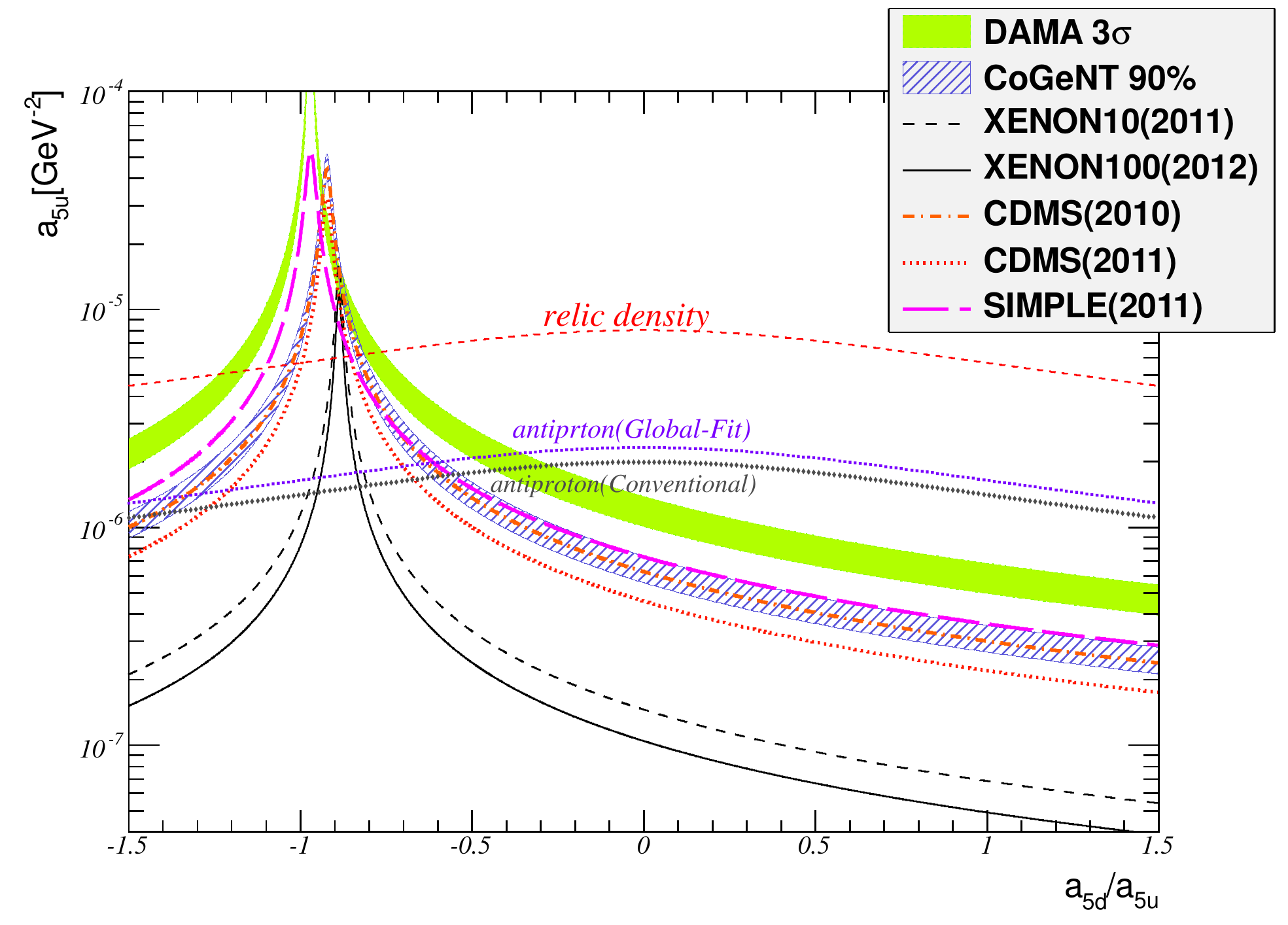}
\end{center}
\caption{
  Upper bounds on the coefficient $a_{5u}$ as a function of $a_{5d}/a_{5u}$ at $95\%$ CL
  from cosmic antiproton flux and DM relic density. The mass of DM particle is fixed at 8 GeV.
  The favored regions and exclusion contours from various experiments such as DAMA \cite{0808.3607},
  GoGeNT \cite{1106.0650}, XENON \cite{1104.3088,1207.5988}, CDMS \cite{1010.4290,1011.2482} and SIMPLE \cite{1106.3014} are also shown.
  }
\label{fig:vector}
\end{figure}

For the operators contribute to $s$-wave annihilation which leads to
velocity-independent annihilation cross sections, the relevant DM couplings to
quarks are found to be tightly constrained by both the cosmic antiproton flux
and the thermal relic density.  In Fig.~\ref{fig:vector} the constraints on
the coefficients $a_{5q}$ for operator
$\mathcal{O}_{5q}=\bar{\chi}\gamma^{\mu}\chi\bar{q}\gamma_{\mu}q$ are shown in
the $(a_{5u}, a_{5d}/a_{5u})$ plane. The mass of the Dirac DM particle is
fixed at $m_{\chi}=8$ GeV.  The results show that the DAMA- and CoGeNT-favored
regions are in tension with both the cosmic antiproton flux and the
thermal relic density.

\begin{figure}[htb]
\begin{center}
  \includegraphics[width=0.9\columnwidth]{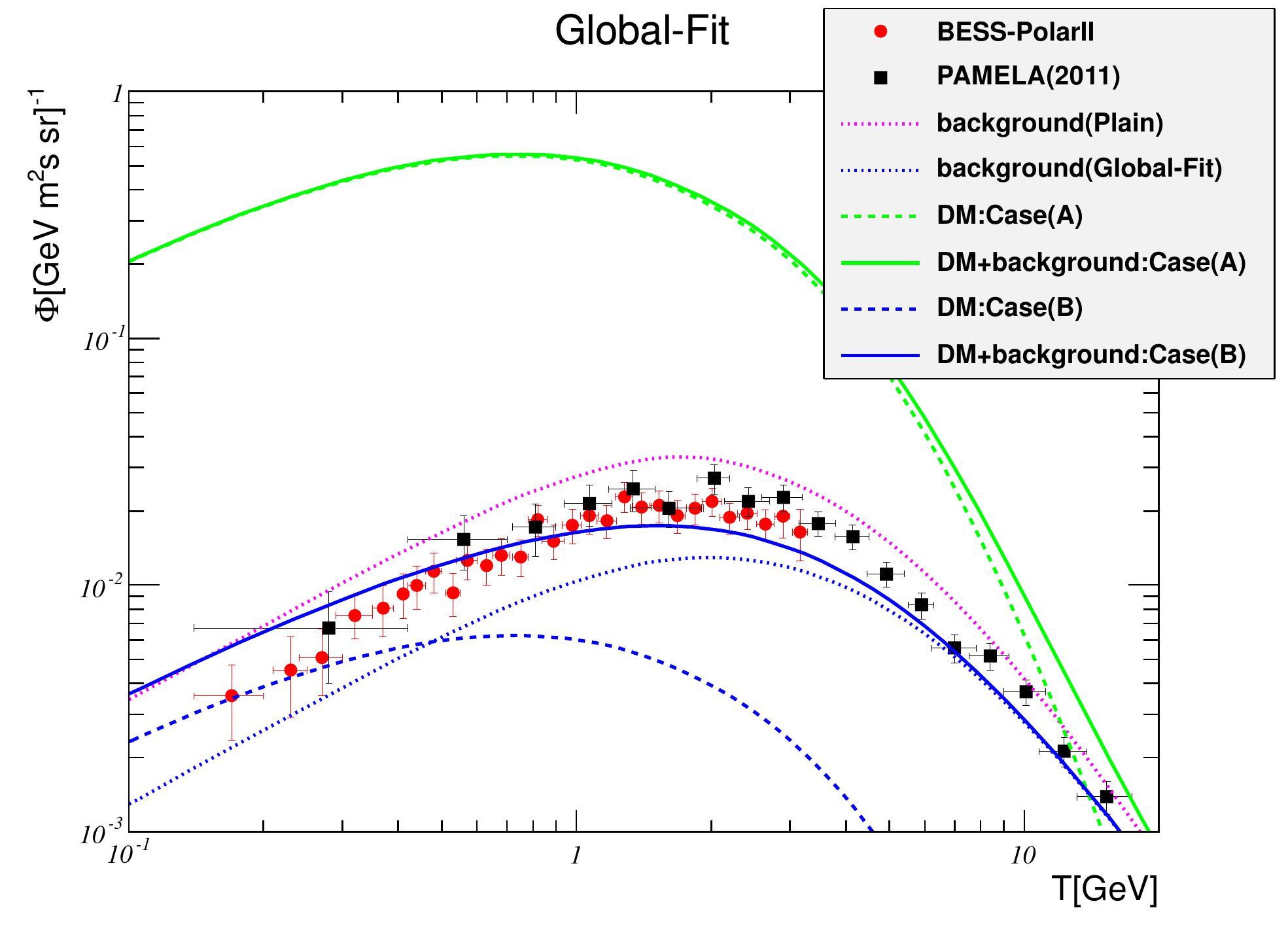}
\end{center}
\caption{ Predictions of cosmic antiproton spectra from DM annihilation
  induced by operator $\mathcal{O}_{5q}$ in the ``Global-Fit'' propagation
  model.  Two cases are considered: (A) For $a_{5u}=1.6\times 10^{-5} \text{
    GeV}^{-2}$ which is favored by the DAMA and CoGeNT experiments. (B) For
  $a_{5u}=1.7\times 10^{-6} \text{ GeV}^{-2}$ which is the maximal value allowed by the
  cosmic antiproton data at $95\%$ CL. The ratio $a_{5d}/a_{5u}$ is fixed at
  $-0.89$ corresponding to $f_n/f_p=-0.70$ and the mass of DM particle is fixed at
  8 GeV. The data of BESS-Polar II \cite{1107.6000} and PAMELA \cite{1103.2880} are also shown;
  right )  Same as left), but  for the ``Conventional'' model. }
\label{fig:vector-pbar}
\end{figure}

At $a_{5d}/a_{5u}=-0.89$ which corresponds to $f_n/f_p=-0.70$, the DAMA and
CoGeNT favored value is $a_{5u} \approx 1.6\times 10^{-5} \text{ GeV}^{-2}$
corresponding to an annihilation cross section of $\langle \sigma v_{\text{rel}} \rangle \approx 3.4\times 10^{-25}
\text{ cm}^{2}$.  As it can be seen in Fig. \ref{fig:vector-pbar}, for such a
large cross section, the predicted antiproton flux is much higher than the
current BESS-Polar II and PAMELA data and results in a  huge
$\chi^2/$d.o.f=$1.3\times 10^{6}/35$  in the ``Global-Fit''  model and an even larger one in the ``Conventional'' model.
The upper bound set by the antiproton data at $95\%$ CL is $a_{5u} \leq 1.7\times 10^{-6} \text{ GeV}^{-2}$ in  the
``Global-Fit '' model at $a_{5d}/a_{5u}=-0.89$, which is about an order of magnitude lower and
corresponds to an annihilation cross section $\langle \sigma v_{\text{rel}} \rangle \approx
3.7\times 10^{-27}  \text{cm}^{2}$. The predicted antiproton fluxes are also shown in Fig.
\ref{fig:vector-pbar}.
In the two  propagation models, the upper bound from the ``Conventional'' model is slightly
stronger than that from the ``Global-Fit'' model.  In
Fig. \ref{fig:vector-pbar}, we also plot the antiproton background of
``Plain'' model which is already higher than the data. Thus the upper bounds
from this propagation model are expected to be  much  stronger. Since we are interested in
conservative upper bounds, we do not further investigate  the constraints in this model.
In Fig. \ref{fig:vector-pbar}, the upper bound from the relic density is
$a_{5u} \leq 6.0\times 10^{-6} \text{ GeV}^{-2}$ at $a_{5d}/a_{5u}=-0.89$,
which is weaker than that from the antiproton flux but still in tension with
the DAMA- and CoGeNT-favored value.

Similar results are found for the operator $\mathcal{O}_{11q}$ \cite{Jin:2012jn}.
Compared with the case of $\mathcal{O}_{5q}$, the constraints on the
coefficients of $\mathcal{O}_{11q}$ are weaker, which is due to the fact that
for the hadronic matrix element of scalar operator $\bar{q}q$ the $B_{iq}$
factors are larger than that for vector operator $\bar{q}\gamma^{\mu}q$, which
allows smaller $a_{iq}$ for the same value of $f_{p,n}$ and results in smaller
annihilation cross sections. The constraint from the thermal relic density is
in tension with the DAMA- and CoGeNT-favored values.
The operators $\mathcal{O}_{1q}= \bar{\chi}\chi\bar{q}q$ and
$\mathcal{O}_{13q}=(\phi^{*}\overleftrightarrow{\partial_{\mu}}\phi )\bar{q}
\gamma^{\mu} q $ contribute to $p$-wave annihilation with cross section
proportional to $v_{\text{rel}}^{2}$. Thus they do contribute very little to
the cosmic antiproton flux. However, their contributions to the thermal relic
density cannot be neglected, as at freeze out the relative velocity is finite.
For the operator $\mathcal{O}_{13q}$ one
can see some tension between bounds set by the relic density and regions
favored by DAMA and CoGeNT data. The constraint is not as stringent as that from the latest
XENON100 data.  For the operator $\mathcal{O}_{1q}$, the
constraints from relic density is rather weak. The difference is again due to
the different $B_{iq}$ factors for these two type of interactions.

It is straight forward to extend the discussions to Majorana fermions and real
scalars. For the particles being its own antiparticles the primary source of
the anitproton will be enhanced by a factor of 2 in Eq. (\ref{eq:ann-source}),
which may lead to more stringent constraints.

In summary, we have investigated the allowed values of DM-nucleon couplings in
the scenario of IVDM for various target nuclei used in DM direct
detections. We find that the recently updated XENON100 result excludes the
main bulk of the overlapping signal region between DAMA and CoGeNT.  We have
shown that whereas the effect of isospin violating scattering can relax the
tensions between the data of DAMA, CoGeNT and XENON, the possible disagreement
between some group of experiments such as that between DAMA and SIMPLE are not
likely to be affected for any value of $f_{n}/f_{p}$.  We have investigated
the conservative constraints on the couplings between the IVDM and the SM
light quarks from the recent cosmic ray antiproton data and that from the
thermal relic density. Among the four operators relevant to IVDM
$\mathcal{O}_{1q}$, $\mathcal{O}_{5q}$, $\mathcal{O}_{11q}$,
$\mathcal{O}_{13q}$, the operators $\mathcal{O}_{5q}$ and $\mathcal{O}_{11q}$
are found to be tightly constrained by the antiproton data and
$\mathcal{O}_{13q}$ is constrained by the relic density. Only the operator
$\mathcal{O}_{1q}$ can survive both the constraints while contribute to large
enough isospin violating interaction required by the current data of DAMA,
CoGeNT and XENON.


This work is supported in part by
the National Basic Research
Program of China (973 Program) under Grants No. 2010CB833000;
the National Nature Science Foundation of China (NSFC) under Grants
No. 10975170,
No. 10821504 and No. 10905084;
and the Project of Knowledge Innovation Program
(PKIP) of the Chinese Academy of Science.





\bibliographystyle{elsarticle-num}
\bibliography{ivpbar,misc,myPub}

\begin{thebibliography}{10}
\expandafter\ifx\csname url\endcsname\relax
  \def\url#1{\texttt{#1}}\fi
\expandafter\ifx\csname urlprefix\endcsname\relax\def\urlprefix{URL }\fi
\expandafter\ifx\csname href\endcsname\relax
  \def\href#1#2{#2} \def\path#1{#1}\fi

\bibitem{0804.2741}
R.~Bernabei, et~al., {First results from DAMA/LIBRA and the combined results
  with DAMA/NaI}, Eur.Phys.J. C56 (2008) 333--355.
\newblock \href {http://arxiv.org/abs/0804.2741} {\path{arXiv:0804.2741}},
  \href {http://dx.doi.org/10.1140/epjc/s10052-008-0662-y}
  {\path{doi:10.1140/epjc/s10052-008-0662-y}}.

\bibitem{0808.3607}
C.~Savage, G.~Gelmini, P.~Gondolo, K.~Freese, {Compatibility of DAMA/LIBRA dark
  matter detection with other searches}, JCAP 0904 (2009) 010.
\newblock \href {http://arxiv.org/abs/0808.3607} {\path{arXiv:0808.3607}},
  \href {http://dx.doi.org/10.1088/1475-7516/2009/04/010}
  {\path{doi:10.1088/1475-7516/2009/04/010}}.

\bibitem{1002.1028}
R.~Bernabei, et~al., {New results from DAMA/LIBRA}, Eur. Phys. J. C67 (2010)
  39--49.
\newblock \href {http://arxiv.org/abs/1002.1028} {\path{arXiv:1002.1028}},
  \href {http://dx.doi.org/10.1140/epjc/s10052-010-1303-9}
  {\path{doi:10.1140/epjc/s10052-010-1303-9}}.

\bibitem{1002.4703}
C.~E. Aalseth, et~al., {Results from a Search for Light-Mass Dark Matter with a
  P- type Point Contact Germanium Detector}, Phys. Rev. Lett. 106 (2011)
  131301.
\newblock \href {http://arxiv.org/abs/1002.4703} {\path{arXiv:1002.4703}},
  \href {http://dx.doi.org/10.1103/PhysRevLett.106.131301}
  {\path{doi:10.1103/PhysRevLett.106.131301}}.

\bibitem{1106.0650}
C.~E. Aalseth, et~al., {Search for an Annual Modulation in a P-type Point
  Contact Germanium Dark Matter Detector}, Phys. Rev. Lett. 107 (2011) 141301.
\newblock \href {http://arxiv.org/abs/1106.0650} {\path{arXiv:1106.0650}},
  \href {http://dx.doi.org/10.1103/PhysRevLett.107.141301}
  {\path{doi:10.1103/PhysRevLett.107.141301}}.

\bibitem{1109.0702}
G.~Angloher, et~al., {Results from 730 kg days of the CRESST-II Dark Matter
  Search}\href {http://arxiv.org/abs/1109.0702} {\path{arXiv:1109.0702}}.

\bibitem{1010.4290}
D.~S. Akerib, et~al., {Low-threshold analysis of CDMS shallow-site data}, Phys.
  Rev. D82 (2010) 122004.
\newblock \href {http://arxiv.org/abs/1010.4290} {\path{arXiv:1010.4290}},
  \href {http://dx.doi.org/10.1103/PhysRevD.82.122004}
  {\path{doi:10.1103/PhysRevD.82.122004}}.

\bibitem{1011.2482}
Z.~Ahmed, et~al., {Results from a Low-Energy Analysis of the CDMS II Germanium
  Data}, Phys. Rev. Lett. 106 (2011) 131302.
\newblock \href {http://arxiv.org/abs/1011.2482} {\path{arXiv:1011.2482}},
  \href {http://dx.doi.org/10.1103/PhysRevLett.106.131302}
  {\path{doi:10.1103/PhysRevLett.106.131302}}.

\bibitem{1104.3088}
J.~Angle, et~al., {A search for light dark matter in XENON10 data}, Phys. Rev.
  Lett. 107 (2011) 051301.
\newblock \href {http://arxiv.org/abs/1104.3088} {\path{arXiv:1104.3088}},
  \href {http://dx.doi.org/10.1103/PhysRevLett.107.051301}
  {\path{doi:10.1103/PhysRevLett.107.051301}}.

\bibitem{1104.2549}
E.~Aprile, et~al., {Dark Matter Results from 100 Live Days of XENON100 Data},
  Phys. Rev. Lett. 107 (2011) 131302.
\newblock \href {http://arxiv.org/abs/1104.2549} {\path{arXiv:1104.2549}},
  \href {http://dx.doi.org/10.1103/PhysRevLett.107.131302}
  {\path{doi:10.1103/PhysRevLett.107.131302}}.

\bibitem{1106.3014}
M.~Felizardo, et~al., {Final Analysis and Results of the Phase II SIMPLE Dark
  Matter Search}\href {http://arxiv.org/abs/1106.3014}
  {\path{arXiv:1106.3014}}.

\bibitem{Guo:2011zze}
W.-L. Guo, Y.-L. Wu, Y.-F. Zhou, {Dark matter candidates in left-right
  symmetric models}, Int.J.Mod.Phys. D20 (2011) 1389--1397.
\newblock \href {http://dx.doi.org/10.1142/S0218271811019578}
  {\path{doi:10.1142/S0218271811019578}}.

\bibitem{Guo:2010sy}
W.-L. Guo, Y.-L. Wu, Y.-F. Zhou, {Searching for Dark Matter Signals in the
  Left-Right Symmetric Gauge Model with CP Symmetry}, Phys.Rev. D82 (2010)
  095004.
\newblock \href {http://arxiv.org/abs/1008.4479} {\path{arXiv:1008.4479}},
  \href {http://dx.doi.org/10.1103/PhysRevD.82.095004}
  {\path{doi:10.1103/PhysRevD.82.095004}}.

\bibitem{Guo:2010vy}
W.-L. Guo, Y.-L. Wu, Y.-F. Zhou, {Exploration of decaying dark matter in a
  left-right symmetric model}, Phys.Rev. D81 (2010) 075014.
\newblock \href {http://arxiv.org/abs/1001.0307} {\path{arXiv:1001.0307}},
  \href {http://dx.doi.org/10.1103/PhysRevD.81.075014}
  {\path{doi:10.1103/PhysRevD.81.075014}}.

\bibitem{Guo:2008si}
W.-L. Guo, L.-M. Wang, Y.-L. Wu, Y.-F. Zhou, C.~Zhuang, {Gauge-singlet dark
  matter in a left-right symmetric model with spontaneous CP violation},
  Phys.Rev. D79 (2009) 055015.
\newblock \href {http://arxiv.org/abs/0811.2556} {\path{arXiv:0811.2556}},
  \href {http://dx.doi.org/10.1103/PhysRevD.79.055015}
  {\path{doi:10.1103/PhysRevD.79.055015}}.

\bibitem{Zhou:2011fr}
Y.-F. Zhou, {Probing the fourth generation Majorana neutrino dark matter},
  Phys.Rev. D85 (2012) 053005.
\newblock \href {http://arxiv.org/abs/1110.2930} {\path{arXiv:1110.2930}},
  \href {http://dx.doi.org/10.1103/PhysRevD.85.053005}
  {\path{doi:10.1103/PhysRevD.85.053005}}.

\bibitem{hep-ph/0307185}
A.~Kurylov, M.~Kamionkowski, {Generalized analysis of weakly-interacting
  massive particle searches}, Phys. Rev. D69 (2004) 063503.
\newblock \href {http://arxiv.org/abs/hep-ph/0307185}
  {\path{arXiv:hep-ph/0307185}}, \href
  {http://dx.doi.org/10.1103/PhysRevD.69.063503}
  {\path{doi:10.1103/PhysRevD.69.063503}}.

\bibitem{hep-ph/0504157}
F.~Giuliani, {Are direct search experiments sensitive to all spin- independent
  WIMP candidates?}, Phys. Rev. Lett. 95 (2005) 101301.
\newblock \href {http://arxiv.org/abs/hep-ph/0504157}
  {\path{arXiv:hep-ph/0504157}}, \href
  {http://dx.doi.org/10.1103/PhysRevLett.95.101301}
  {\path{doi:10.1103/PhysRevLett.95.101301}}.

\bibitem{1003.0014}
A.~L. Fitzpatrick, D.~Hooper, K.~M. Zurek, {Implications of CoGeNT and DAMA for
  Light WIMP Dark Matter}, Phys. Rev. D81 (2010) 115005.
\newblock \href {http://arxiv.org/abs/1003.0014} {\path{arXiv:1003.0014}},
  \href {http://dx.doi.org/10.1103/PhysRevD.81.115005}
  {\path{doi:10.1103/PhysRevD.81.115005}}.

\bibitem{1004.0697}
S.~Chang, J.~Liu, A.~Pierce, N.~Weiner, I.~Yavin, {CoGeNT Interpretations},
  JCAP 1008 (2010) 018.
\newblock \href {http://arxiv.org/abs/1004.0697} {\path{arXiv:1004.0697}},
  \href {http://dx.doi.org/10.1088/1475-7516/2010/08/018}
  {\path{doi:10.1088/1475-7516/2010/08/018}}.

\bibitem{1102.4331}
J.~L. Feng, J.~Kumar, D.~Marfatia, D.~Sanford, {Isospin-Violating Dark Matter},
  Phys. Lett. B703 (2011) 124--127.
\newblock \href {http://arxiv.org/abs/1102.4331} {\path{arXiv:1102.4331}},
  \href {http://dx.doi.org/10.1016/j.physletb.2011.07.083}
  {\path{doi:10.1016/j.physletb.2011.07.083}}.

\bibitem{1105.3734}
M.~T. Frandsen, et~al., {On the DAMA and CoGeNT Modulations}, Phys. Rev. D84
  (2011) 041301.
\newblock \href {http://arxiv.org/abs/1105.3734} {\path{arXiv:1105.3734}},
  \href {http://dx.doi.org/10.1103/PhysRevD.84.041301}
  {\path{doi:10.1103/PhysRevD.84.041301}}.

\bibitem{1112.6364}
X.-G. He, B.~Ren, J.~Tandean, {Hints of Standard Model Higgs Boson at the LHC
  and Light Dark Matter Searches}, Phys.Rev. D85 (2012) 093019.
\newblock \href {http://arxiv.org/abs/1112.6364} {\path{arXiv:1112.6364}},
  \href {http://dx.doi.org/10.1103/PhysRevD.85.093019,
  10.1103/PhysRevD.85.119902, 10.1103/PhysRevD.85.119906}
  {\path{doi:10.1103/PhysRevD.85.093019, 10.1103/PhysRevD.85.119902,
  10.1103/PhysRevD.85.119906}}.

\bibitem{1103.3270}
J.~Kumar, J.~G. Learned, M.~Sakai, S.~Smith, {Dark Matter Detection With
  Electron Neutrinos in Liquid Scintillation Detectors}, Phys.Rev. D84 (2011)
  036007.
\newblock \href {http://arxiv.org/abs/1103.3270} {\path{arXiv:1103.3270}},
  \href {http://dx.doi.org/10.1103/PhysRevD.84.036007}
  {\path{doi:10.1103/PhysRevD.84.036007}}.

\bibitem{1106.4044}
S.-L. Chen, Y.~Zhang, {Isospin-Violating Dark Matter and Neutrinos From the
  Sun}, Phys.Rev. D84 (2011) 031301.
\newblock \href {http://arxiv.org/abs/1106.4044} {\path{arXiv:1106.4044}},
  \href {http://dx.doi.org/10.1103/PhysRevD.84.031301}
  {\path{doi:10.1103/PhysRevD.84.031301}}.

\bibitem{1112.4849}
J.~Kumar, D.~Sanford, L.~E. Strigari, {New Constraints on Isospin-Violating
  Dark Matter}, Phys.Rev. D85 (2012) 081301.
\newblock \href {http://arxiv.org/abs/1112.4849} {\path{arXiv:1112.4849}}.

\bibitem{1107.6000}
K.~Abe, et~al., {Measurement of the cosmic-ray antiproton spectrum at solar
  minimum with a long-duration balloon flight over Antarctica}, Phys. Rev.
  Lett. 108 (2012) 051102.
\newblock \href {http://arxiv.org/abs/1107.6000} {\path{arXiv:1107.6000}},
  \href {http://dx.doi.org/10.1103/PhysRevLett.108.051102}
  {\path{doi:10.1103/PhysRevLett.108.051102}}.

\bibitem{1103.2880}
O.~Adriani, et~al., {The cosmic-ray electron flux measured by the PAMELA
  experiment between 1 and 625 GeV}, Phys.Rev.Lett. 106 (2011) 201101.
\newblock \href {http://arxiv.org/abs/1103.2880} {\path{arXiv:1103.2880}},
  \href {http://dx.doi.org/10.1103/PhysRevLett.106.201101}
  {\path{doi:10.1103/PhysRevLett.106.201101}}.

\bibitem{Jin:2012jn}
H.-B. Jin, S.~Miao, Y.-F. Zhou, {Implications of the latest XENON100 and cosmic
  ray antiproton data for isospin violating dark matter}\href
  {http://arxiv.org/abs/1207.4408} {\path{arXiv:1207.4408}}.

\bibitem{1207.5988}
E.~Aprile, et~al., {Dark Matter Results from 225 Live Days of XENON100
  Data}\href {http://arxiv.org/abs/1207.5988} {\path{arXiv:1207.5988}}.

\bibitem{1106.3559}
J.~Collar, {Comments on 'Final Analysis and Results of the Phase II SIMPLE Dark
  Matter Search'}\href {http://arxiv.org/abs/1106.3559}
  {\path{arXiv:1106.3559}}.

\bibitem{1107.1515}
{Reply to arxiv:1106.3559 by J.I. Collar}\href {http://arxiv.org/abs/1107.1515}
  {\path{arXiv:1107.1515}}.

\bibitem{1103.3481}
J.~Collar, {A comparison between the low-energy spectra from CoGeNT and
  CDMS}\href {http://arxiv.org/abs/1103.3481} {\path{arXiv:1103.3481}}.

\bibitem{1005.0838}
J.~Collar, D.~McKinsey, {Comments on 'First Dark Matter Results from the
  XENON100 Experiment'}\href {http://arxiv.org/abs/1005.0838}
  {\path{arXiv:1005.0838}}.

\bibitem{1005.2615}
{Reply to the Comments on the XENON100 First Dark Matter Results}\href
  {http://arxiv.org/abs/1005.2615} {\path{arXiv:1005.2615}}.

\bibitem{1110.5338}
C.~Kelso, D.~Hooper, M.~R. Buckley, {Toward A Consistent Picture For CRESST,
  CoGeNT and DAMA}, Phys.Rev. D85 (2012) 043515.
\newblock \href {http://arxiv.org/abs/1110.5338} {\path{arXiv:1110.5338}},
  \href {http://dx.doi.org/10.1103/PhysRevD.85.043515}
  {\path{doi:10.1103/PhysRevD.85.043515}}.

\bibitem{astro-ph/9807150}
A.~Strong, I.~Moskalenko, {Propagation of cosmic-ray nucleons in the galaxy},
  Astrophys.J. 509 (1998) 212--228.
\newblock \href {http://arxiv.org/abs/astro-ph/9807150}
  {\path{arXiv:astro-ph/9807150}}, \href {http://dx.doi.org/10.1086/306470}
  {\path{doi:10.1086/306470}}.

\bibitem{astro-ph/0106567}
I.~V. Moskalenko, A.~W. Strong, J.~F. Ormes, M.~S. Potgieter, {Secondary
  anti-protons and propagation of cosmic rays in the galaxy and heliosphere},
  Astrophys.J. 565 (2002) 280--296.
\newblock \href {http://arxiv.org/abs/astro-ph/0106567}
  {\path{arXiv:astro-ph/0106567}}, \href {http://dx.doi.org/10.1086/324402}
  {\path{doi:10.1086/324402}}.

\bibitem{astro-ph/0101068}
A.~Strong, I.~Moskalenko, {Models for galactic cosmic ray propagation},
  Adv.Space Res. 27 (2001) 717--726.
\newblock \href {http://arxiv.org/abs/astro-ph/0101068}
  {\path{arXiv:astro-ph/0101068}}, \href
  {http://dx.doi.org/10.1016/S0273-1177(01)00112-0}
  {\path{doi:10.1016/S0273-1177(01)00112-0}}.

\bibitem{astro-ph/0210480}
I.~V. Moskalenko, A.~Strong, S.~Mashnik, J.~Ormes, {Challenging cosmic ray
  propagation with antiprotons. Evidence for a fresh nuclei component?},
  Astrophys.J. 586 (2003) 1050--1066.
\newblock \href {http://arxiv.org/abs/astro-ph/0210480}
  {\path{arXiv:astro-ph/0210480}}, \href {http://dx.doi.org/10.1086/367697}
  {\path{doi:10.1086/367697}}.

\bibitem{astro-ph/0510335}
V.~Ptuskin, I.~V. Moskalenko, F.~Jones, A.~Strong, V.~Zirakashvili,
  {Dissipation of magnetohydrodynamic waves on energetic particles: impact on
  interstellar turbulence and cosmic ray transport}, Astrophys.J. 642 (2006)
  902--916.
\newblock \href {http://arxiv.org/abs/astro-ph/0510335}
  {\path{arXiv:astro-ph/0510335}}, \href {http://dx.doi.org/10.1086/501117}
  {\path{doi:10.1086/501117}}.

\bibitem{1001.0551}
A.~Putze, L.~Derome, D.~Maurin, {A Markov Chain Monte Carlo technique to sample
  transport and source parameters of Galactic cosmic rays: II. Results for the
  diffusion model combining B/C and radioactive nuclei}, Astron.Astrophys. 516
  (2010) A66.
\newblock \href {http://arxiv.org/abs/1001.0551} {\path{arXiv:1001.0551}}.

\bibitem{1011.0037}
R.~Trotta, G.~Johannesson, I.~Moskalenko, T.~Porter, R.~R. de~Austri, et~al.,
  {Constraints on cosmic-ray propagation models from a global Bayesian
  analysis}, Astrophys.J. 729 (2011) 106.
\newblock \href {http://arxiv.org/abs/1011.0037} {\path{arXiv:1011.0037}},
  \href {http://dx.doi.org/10.1088/0004-637X/729/2/106}
  {\path{doi:10.1088/0004-637X/729/2/106}}.

\bibitem{astro-ph/9712318}
L.~Bergstrom, P.~Ullio, J.~H. Buckley, {Observability of gamma-rays from dark
  matter neutralino annihilations in the Milky Way halo}, Astropart.Phys. 9
  (1998) 137--162.
\newblock \href {http://arxiv.org/abs/astro-ph/9712318}
  {\path{arXiv:astro-ph/9712318}}, \href
  {http://dx.doi.org/10.1016/S0927-6505(98)00015-2}
  {\path{doi:10.1016/S0927-6505(98)00015-2}}.

\bibitem{1001.4538}
E.~Komatsu, et~al., {Seven-Year Wilkinson Microwave Anisotropy Probe (WMAP)
  Observations: Cosmological Interpretation}, Astrophys.J.Suppl. 192 (2011) 18.
\newblock \href {http://arxiv.org/abs/1001.4538} {\path{arXiv:1001.4538}},
  \href {http://dx.doi.org/10.1088/0067-0049/192/2/18}
  {\path{doi:10.1088/0067-0049/192/2/18}}.

\end{thebibliography}






\end{document}